\documentclass[10pt, journal, onecolumn, letterpaper, twoside]{IEEEtran}%

\usepackage{amssymb}
\usepackage{amsmath}
\usepackage{bm}
\usepackage{amsbsy}
\usepackage[mathcal]{eucal}
\usepackage{cite}
\usepackage{graphicx}
\usepackage{subfigure}
\usepackage{amsfonts}%
\providecommand{\U}[1]{\protect\rule{.1in}{.1in}}
\begin{document}

\title{An Approximation of the First Order Marcum $Q$-Function with Application to Network Connectivity Analysis}
\author{Mohammud Z. Bocus, Carl P. Dettmann and Justin P. Coon\thanks{M. Z. Bocus and J. P. Coon are with the
Telecommunications Research Laboratory, Toshiba Research Europe Ltd., 32 Queen
Square, Bristol, BS1 4ND, U.K.; tel: +44 (0)117 906 0700, fax: +44 (0)117 906
0701. (e-mail: zubeir.bocus@toshiba-trel.com).}
\thanks{C. P. Dettmann is with the University of Bristol School of
Mathematics, University Walk, Bristol, UK, BS8 1TW.}
\thanks{J. P. Coon is also with the Department of
Electrical and Electronic Engineering, University of Bristol, BS8 1UB, U.K.}
\thanks{\copyright 2013 IEEE. Personal use of this material is permitted. Permission from IEEE must be obtained for all other uses, in any current or future media, including reprinting/republishing this material for advertising or promotional purposes, creating new collective works, for resale or redistribution to servers or lists, or reuse of any copyrighted component of this work in other works.}}
\maketitle

\begin{abstract}
An exponential-type approximation of the first order Marcum $Q$-function is presented, which is robust to changes in its first argument and can easily be integrated with respect to the second argument. Such characteristics are particularly useful in network connectivity analysis. The proposed approximation is exact in the limit of small first argument of the Marcum $Q$-function, in which case the optimal parameters can be obtained analytically. For larger values of the first argument, an optimization problem is solved, and the parameters can be accurately represented using regression analysis. Numerical results indicate that the proposed methods result in approximations very close to the actual Marcum $Q$-function for small and moderate values of the first argument. We demonstrate the accuracy of the approximation by using it to analyze the connectivity properties of random ad hoc networks operating in a Rician fading environment.
\end{abstract}

\section{Introduction}

The Marcum $Q$-function, defined as the integral~\cite{Simon2003}
\begin{equation}\label{eq:marcumq}
Q_{1}(a,b) = \int_{b}^{\infty} x \exp\left(  -\frac{x^{2}+a^{2}}{2}\right)
I_{0}(ax) \mathrm{d}x
\end{equation}
for $a,b\geq0$ where $I_{0}(x)$ is the modified Bessel function of the first kind, is a
fundamental function that arises in the performance evaluation of a wide class
of communication systems \cite{Simon2003, Odriscoll2009a, Fu2011}.
From a mathematical point of view, this function represents the
complementary cumulative distribution function (CCDF) of the power of a Rician
distribution. The
integral representation of the function given by~\eqref{eq:marcumq} cannot be 
manipulated easily to provide simple expressions for the performance of communication systems, especially when the function $Q_1(a,b)$ must be integrated
with respect to one of its arguments \cite{Simon2003}. To solve this issue,
numerous works have proposed alternative representations of $Q_1(a,b)$
to facilitate analysis (see, e.g.,~\cite{Zhao2008, Fu2011} and references therein).
Exponential-type bounds, provided they are tight, have been particularly
attractive, especially when evaluating the bit error rate at high signal to noise ratio (SNR) \cite{Simon2000,
Fu2011}.  In other situations, approximations may be more suitable than bounds \cite{Ding2008, Sofotasios2010a}.  However, such approximations may have complicated mathematical structures and/or be inaccurate in certain domains of their arguments.

In this paper, a simple exponential approximation of the first order Marcum $Q$-function is
presented that yields small approximation error over a large domain in its two arguments.  
The approximation is designed such that it can be used in situations where $Q_1(a,b)$ must be integrated over its second argument.
In what follows, a heuristic approach is first employed to find the right form of the
approximation, which is parameterized by two functions of $a$.  An analytical framework for determining the correct parameterization is then explored, which is shown to be accurate for $0 \leq a \ll 1$.  For $a \gg 1$, the optimal parameterization is calculated numerically.  Although the proposed approximation is useful in its own right, it is particularly helpful in situations where the
$Q$-function must be integrated. 
To illustrate this fact, we present an
example application of our results whereby the proposed approximation is used
to analyze the connectivity probability of a random ad hoc network operating
in Rician fading channels.

The structure of this paper is as follows. In Section
\ref{sec:proposed_approx}, the proposed approximation and means of deriving
the optimal $a$-dependent parameters are presented. An example application of the approximation is given in Section \ref{sec:application},
while the accuracy of the approximation is presented in Section
\ref{sec:results}. Finally, some concluding remarks are given in Section
\ref{sec:conclusion}.

\section{Approximation of $Q_{1}(a,b)$}

\label{sec:proposed_approx} By plotting $Q_1(a,b)$ as a function of $b$ for various values of $a$, it can be readily observed that $Q_{1}(a,b)$ decays exponentially with $b$, where the value of $a$ roughly defines the shift of $Q_1$ along the $b$-axis.
Consequently, we propose to approximate $Q_{1}(a,b)$ by the function%
\begin{equation}
\tilde{Q}_{1}(a,b)=\exp\left(  -e^{\nu(a)}b^{\mu(a)}\right)  \label{eq:approx}%
\end{equation}
where $\nu(a)$ and $\mu(a)$ are nonnegative parameters dependent upon $a$. The
key is to choose these parameters such that the accuracy of the approximation
is high.

As previously discussed, we are concerned with obtaining an approximation that
is useful over the range of the argument $b$ for some fixed $a$. Thus, we
define the approximation error as the function%
\begin{equation}
\mathcal{E}(a)=\int_{0}^{\infty}(Q_{1}(a,b)-\tilde{Q}_{1}(a,b))^{2}%
\mathrm{d}b.\label{eq:approx_error}%
\end{equation}
Furthermore, we define $\mu(a)$ and $\nu(a)$ to be polynomials of order $m$,
where a larger value of $m$ yields a better approximation. Thus, we have%
\begin{align*}
\mu(a) &  =\mu_{0}+\mu_{1}a+\mu_{2}a^{2}+\cdots+\mu_{m}a^{m}\\
\nu(a) &  =\nu_{0}+\nu_{1}a+\nu_{2}a^{2}+\cdots+\nu_{m}a^{m}%
\end{align*}
in which case the approximation becomes%
\begin{equation}
\tilde{Q}_{1}(a,b)=\exp\left(  -e^{\sum_{n=0}^{m}(\mu_{n}\ln b+\nu_{n})a^{n}%
}\right)  .\label{eq:approx_expand}%
\end{equation}
The goal is now to choose the coefficients $\{\mu_{0},\ldots,\mu_{m},\nu
_{0},\ldots,\nu_{m}\}$, independent of $b$, such that $\mathcal{E}(a)$ is minimized. Depending on
the value of the argument $a$, this can be done analytically or numerically.

\subsection{Analytical Approach for Small Arguments}

First, consider the case where $0\leq a\ll1$. It is logical to expand $Q_{1}(a,b)$
and $\tilde{Q}_{1}(a,b)$ about $a=0$ and equate the coefficients term by term.
Of course, if the expansions for $Q$ and $\tilde{Q}$ converge and the
corresponding coefficients match to arbitrary order, then $Q=\tilde{Q}$. Since
this is clearly not the case, it is advisable to equate coefficients
recursively, from lowest to highest order.

\begin{figure}[t]
\centering
\includegraphics[width=7.5cm]{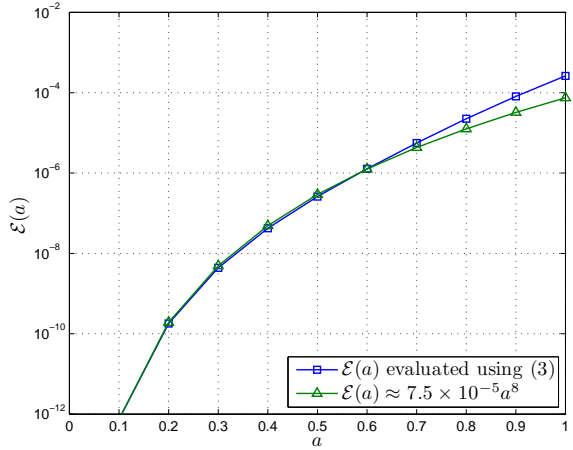}\caption{Approximation error derived from fourth order polynomial representation of $\mu(a)$ and $\nu(a)$ for small $a$.}%
\label{fig:approximation_error_small_a}%
\end{figure}

For example, let $m=4$. To leading order, we have $Q_{1}(0,b)=\exp(-b^{2}/2)$
and $\tilde{Q}_{1}(0,b)=\exp(-e^{\nu_{0}}b^{\mu_{0}})$. It follows that we
should choose $\mu_{0}=2$ and $\nu_{0}=-\ln2$ since this ensures the
approximation is exact at $a=0$. Next, we can equate the first order terms to
obtain the equation\footnote{The details of the calculations are
straightforward but lengthy, and are thus omitted here for brevity.} $\mu
_{1}\ln b+\nu_{1}=0$. But we see from (\ref{eq:approx_expand}) that this
formula implies there is no $O(a)$ term in the second exponent of $\tilde{Q}$.
Thus, we may take $\mu_{1}=\nu_{1}=0$ to maintain independence of $b$. This
process can be continued in a straightforward manner. However, when we equate
the fourth order terms, we obtain the equation $\mu_{4}\ln b+\nu_{4}=b^{2}%
/32$, and thus either $\mu_4$ or $\nu_4$ is
dependent upon $b$, a condition that is not allowed by our
definition of the polynomials $\mu$ and $\nu$. Instead, we can optimize $\mathcal{E}(a)$ over $\mu_{4}$ and
$\nu_{4}$ by differentiating with respect to each variable, setting the
results to zero, and solving for $\mu_{4}$ and $\nu_{4}$. This yields
the optimal fourth order polynomials%
\begin{align*}
\mu(a) &  =2+\frac{9}{8(9\pi^{2}-80)}a^{4}\\
\nu(a) &  =-\ln2-\frac{a^{2}}{2}+\frac{45\pi^{2}+72\ln2+36C-496}{64(9\pi
^{2}-80)}a^{4}%
\end{align*}
which are independent of $b$, and thus satisfy the conditions of our
approximation\footnote{The symbol $C$ denotes the Euler-Mascheroni constant,
where $C\approx0.5772$.}. By substituting these expressions for $\mu(a)$ and
$\nu(a)$ into $\tilde{Q}$ and evaluating the integral in
(\ref{eq:approx_error}) for small $a$, it is apparent that $\mathcal{E}(a)\approx
7.5\times10^{-5}a^{8}$. Thus, the fourth order result is very accurate for
$a\ll1$, an observation that is corroborated by Fig.~\ref{fig:approximation_error_small_a}.

\subsection{Numerical Approach for General Arguments}
While a closed-form expression for the coefficients of $\mu$ and $\nu$ can be obtained for small $a$, performing a similar analysis for larger values of $a$ is somewhat problematic.   On that account, a numerical approach is followed instead.  In particular, we propose to determine the appropriate values of $\mu$ and $\nu$ such that the following error is minimized:
\begin{equation}
\hat{\mathcal{E}} = \delta\sum_{\beta=0}^{\infty}\left( Q_{1}(a,\delta\beta)-\exp\left(-e^{\nu}(\delta\beta)^{\mu}\right) \right) ^{2} 
\label{eq:SSE_sum}
\end{equation}
where $\delta$ is small and $\hat{\mathcal{E}} \rightarrow \mathcal{E}$ as $\delta\rightarrow 0$. Such optimization problems can be solved using numerical techniques \cite{Bjorck1996}. It should be noted that the problem of minimizing the error term defined in (\ref{eq:SSE_sum}) is not a convex optimization problem. As such, numerical methods may not always converge to the global optimum. Nevertheless, we find that the observed optimum is often adequate, as illustrated in Section \ref{sec:results}.

Since $Q_{1}(a,b)$ decays exponentially with $b$, we argue that we can ignore terms in the summation in~\eqref{eq:SSE_sum} corresponding to values of $b$ larger than some $b_{\max}$ in order to facilitate optimization.  This is particularly justified by noting that we are interested in obtaining an accurate
expression for $Q_{1}$ that captures most of its mass. Thus, the upper limit
on the summation in (\ref{eq:SSE_sum}) can be replaced by $\beta_{\max} =
b_{\max}/\delta$. As an example, we set $\delta=10^{-4}$ and $b_{\max}=12$, and solved the above optimization problem using a line-search algorithm for several values of $a$. Results are shown in Table \ref{tab:nu_tab}.

\begin{table}[t]
\caption{Suitable $\nu$ and $\mu$ for different $a$}%
\centering
\label{tab:nu_tab}
\begin{tabular}[c]{ccc}\hline\hline
$a$ & $\nu$ & $\mu$\\[0.5ex]\hline
$1.0000$ & $-1.1739$ & $2.0921$\\
$2.0000$ & $-2.5492$ & $2.7094$\\
$3.0000$ & $-4.6291$ & $3.6888$\\
$4.0000$ & $-7.1668$ & $4.7779$\\
$5.0000$ & $-10.0339$ & $5.9074$\\
$6.0000$ & $-13.2014$ & $7.0794$ \\[1ex]\hline
\end{tabular}
\end{table}

Using the values listed in Table~\ref{tab:nu_tab}, it is possible to derive an approximate expression for $\mu(a)$ and $\nu(a)$ using polynomial regression \cite{Luxmoore1979}.  For instance, assuming that $\mu(a)$ is a polynomial of fourth order in $a$, the regression model for $\mu(a)$ can be expressed as
\begin{equation}
\mu(a_j) = \tilde{\mu}_0 + \tilde{\mu}_1 a_j + \tilde{\mu}_2 a^2_j + \tilde{\mu}_3 a^3_j +  \tilde{\mu}_4 a^4_j + \epsilon_j
\end{equation}
for $j = 1,\ldots,N$ where $\epsilon_j$ is the error in the approximation, $\{\tilde{\mu}_i\}_{i=1}^4$ are the estimation coefficients, and $N$ is the number of observed instances (c.f.~Table~\ref{tab:nu_tab}).  The above expression can be written in matrix form as $\boldsymbol{\mu} = \mathbf{A} \tilde{\boldsymbol{\mu}} + \boldsymbol{\epsilon}$, where $\tilde{\boldsymbol{\mu}} = \left[\tilde{\mu}_0,\cdots,\tilde{\mu}_4 \right]^T$ and $\boldsymbol{\mu}$ is similarly defined, $\mathbf{A}$ is an $N\times 5$ matrix with $k$th column being $\left[a_1^{k-1},\cdots, a_N^{k-1} \right]^T $ for $k=1,\ldots,5$, and $\boldsymbol{\epsilon} = \left[{\epsilon}_1,\cdots,\tilde{\epsilon}_N \right]^T$.  Using ordinary least squares estimation, the coefficients can be obtained using
$\tilde{\boldsymbol{\mu}} = \left(\mathbf{A}^T \mathbf{A} \right)^{-1}\mathbf{A}^T \boldsymbol{\mu}$.  For $m=4$, this approach yields
\begin{align}
\mu(a)  &  = 2.174 -0.592a +0.593a^{2} -0.092a^{3} +0.005a^{4} \nonumber \\
\nu(a)  & = -0.840  + 0.327 a -0.740 a^2 + 0.083 a^3 - 0.004 a^4.
\label{eq:approx_mu}%
\end{align}

\begin{figure}[t]
\centering
\includegraphics[width=7.5cm]{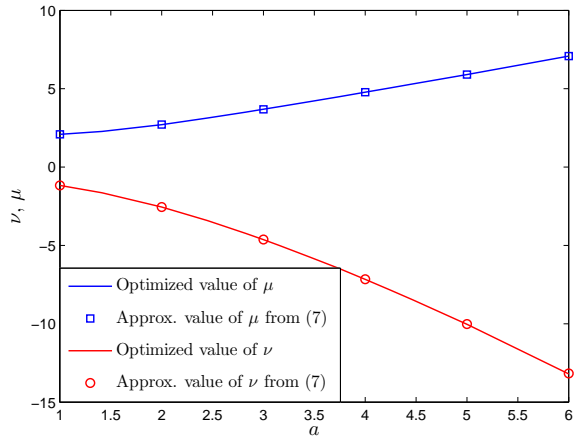}
\caption{Plots of $\mu(a)$ and $\nu(a)$ obtained through a numerical approach. The solid lines represent the optimized values, while the squares and circles
show the approximate values from (\ref{eq:approx_mu}).}%
\label{fig:mu_nu_a_plot}%
\end{figure}

Fig. \ref{fig:mu_nu_a_plot} depicts the comparison between
the optimized values (from Table \ref{tab:nu_tab}) and approximated values of
the two parameters given in (\ref{eq:approx_mu}). It can be observed from the plots that the two sets of
values are very close, indicating the suitability of the above two equations.
Such approximations are convenient if fast computation of the parameters
$\nu(a)$ and $\mu(a)$ are required. 

\section{An Application of the Proposed Approximation}
\label{sec:application} 

As previously stated, the presented approximation can
be particularly useful when the CCDF of the power of a Rician
channel needs to be
integrated over the second argument. The power of a Rician channel is noncentral-$\chi^{2}$
distributed, whose CCDF is given by
\begin{equation}
F_{X}(x) = Q_{1}\Big(  \sqrt{2K},\sqrt{{2\omega^{-1}(K+1)x}} \Big)
\end{equation}
where $K$ is the Rice factor and $\omega$ is a channel dependent parameter.
Given that $a = \sqrt{2K}$ in this case and, in general, $1\leq K \leq10$
\cite{Proakis2000}, it follows that $a< 5$. On that account, the
approximations of the parameters $\mu(a)$ and $\nu(a)$ presented above can
readily be used.

To demonstrate the use of the proposed approximation of the Marcum
$Q$-function, we consider the analysis of the full connection probability of a
random ad hoc network, similar to the work presented in \cite{Miorandi2008,
Coon2012a}. Consider the connection probability of two nodes in a system,
which we denote by $H$, given a minimum data rate requirement of $R_{0}$. By
adopting an information theoretic definition of connectivity, we define
\begin{equation}
H = P\left(  \log_{2}(1+\gamma|h|^{2}) \geq R_{0} \right)
\end{equation}
where $|h|^{2}$ is the channel gain between the two nodes and $\gamma$ is the
SNR which is dependent upon the distance between the two nodes and other
parameters such as the path loss exponent and antenna gains. Under the
assumption of a Rician fading channel with Rice factor $K$ and a path loss
exponent of two for illustration, we have $H(r) = Q_{1}\big(  \sqrt{2K}, r\alpha\big)$ 
where $r$ is the distance between the two nodes and $\alpha$ is a function of
the system parameters. To derive the probability that the
network is fully connected, it is necessary
to average $H(r)$ over the configuration space \cite{Coon2012a}. For a
homogeneous system, this amounts to averaging $H(r)$ over all distances
between nodes. Such a calculation would involve an integral of the form (c.f.,
(19)-(21) in \cite{Coon2012a} for Rayleigh fading)
\begin{align}
\int_{r_{1}}^{r_{2}} r H(r) dr  &  = \int_{r_{1}}^{r_{2}} r Q_{1}\Big(
\sqrt{2K}, r\alpha\Big)  dr \nonumber\\
& \approx\int_{r_{1}}^{r_{2}} r e^{-e^{\nu}(r\alpha)^{\mu}} dr\nonumber\\
&  = \frac{1}{\mu}\lambda^{-\frac{2}{\mu}} \left(  \gamma(\frac{2}{\mu
},\lambda r_{2}^{\mu}) - \gamma(\frac{2}{\mu},\lambda r_{1}^{\mu}) \right)
\end{align}
where $\lambda = e^{\nu}\alpha^{\mu}$, $r_{1}$ and $r_{2}$ are the minimum and maximum distances between nodes
within the system, and $\gamma(x,y)$ is the lower incomplete gamma function.
The complete analysis of the full connection probability is beyond the scope
of this letter. What is important to note is that without the approximation
derived in this paper, solving the integral stated above would be very
challenging if not impossible.

\section{Accuracy of the Approximation}
\label{sec:results} 

\begin{figure}[t]
	\centering
		\includegraphics[width=8cm]{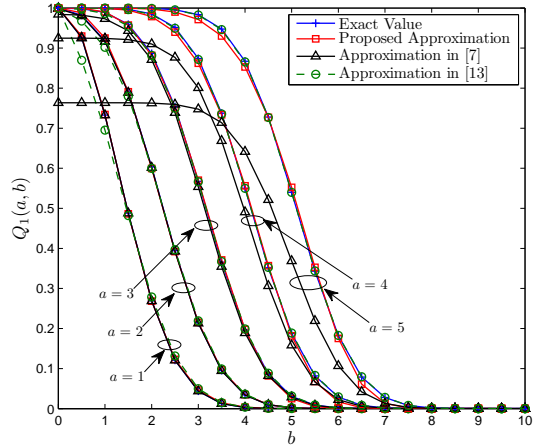}
	\caption{Comparison of the Marcum $Q$-function with the proposed approximation and that presented in \cite{Sofotasios2010a} and \cite{Li2010}. For larger $a$, the approximation in \cite{Sofotasios2010a} diverges from the actual curve for small $b$.  Parameters for the proposed approximations are obtained from (\ref{eq:approx_mu}).}
	\label{fig:comparison_apprx}
\end{figure}

To demonstrate the accuracy of the proposed approximation, we compare the proposed method to existing approximations in the literature \cite{Sofotasios2010a, Li2010} that generally yield small approximation errors.  The comparisons are shown in Fig. \ref{fig:comparison_apprx}.  For the approximation presented in \cite{Sofotasios2010a}, the value of $k$ was set to $50$ in equation (6) therein.  On the other hand, the approximation in \cite{Li2010} is obtained by taking the average of the lower and upper bounds of the $Q-$function as presented by the authors.  It can be observed from the plot that, for small $a$, the approximations are close to the Marcum $Q$-function.  However, as $a$ increases, divergence from the actual curve is seen for the approximation of \cite{Sofotasios2010a}.  Nevertheless, the proposed approximation still adequately represents the mass of the Marcum $Q$-function over the range of $b$ values; consequently, our approximation is robust with respect to changes in $a$, similar to \cite{Li2010}.  It should be noted that, although the integral of the approximation in \cite{Li2010} is possible, the resulting mathematical expressions are considerably more complicated than the proposed method and thus do not easily lend themselves to further manipulations and calculations.

For large values of $b$, we note that existing bounds and approximations in the literature often provide a more accurate representation of $Q_1(a,b)$ compared to the proposed method.  This can easily be observed graphically, but we omit the results here due to space constraints.  We would also like to point out that some existing approximations \cite{Pent1968} lead to very accurate representations of $Q_1(a,b)$ for all $a$ and small values of $b$.  As $b$ increases however, such approximations diverge.

For applications that require a closer approximation for large $b$, the expressions in \cite{Sofotasios2010a, Zhao2008} and references therein would be more appropriate.  However, if the integral of the Marcum $Q$-function over the domain of the second argument is sought, the approximation presented in this paper is more suitable.

\begin{figure}[t]
\centering
\includegraphics[width=8cm]{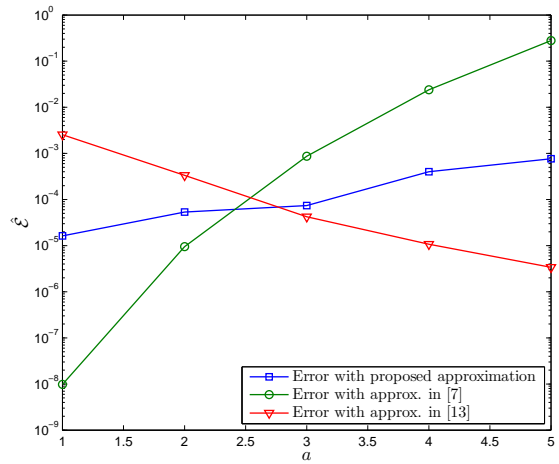}\caption{Comparison
of the approximation error $\hat{\mathcal{E}}$ using the proposed approach and the ones given in \cite{Sofotasios2010a} and \cite{Li2010}.  The proposed approximation remains robust to changes in $a$.
}%
\label{fig:SSE_plot}%
\end{figure}

We next compare $\hat{\mathcal{E}}$ for our proposed approximation and the one given in~\cite{Sofotasios2010a,Li2010}.  Results shown in Fig. \ref{fig:SSE_plot} for different values of $a$ demonstrate the accuracy of the proposed scheme.  As mentioned in the previous section, the range of $a$ values considered in the plot is the range that would typically be encountered in practice in communication system analysis with Rician fading \cite{Proakis2000}.  However, for the problem defined in (\ref{eq:SSE_sum}), it is guaranteed that the solution would minimize the error term for any value of $a$.  

\section{Conclusion}

\label{sec:conclusion} 
In this paper, a simple 
approximation of the first order Marcum $Q$-function was presented that can
be used in network connectivity analysis.  For small input argument $a$, an analytical approach was 
presented for finding the approximation parameters, while for larger $a$, 
a numerical procedure based on an optimization problem was proposed.  
Equations for approximating these parameters were then presented. Simulation results demonstrated that the
approximations led to an accurate representation of the Marcum $Q$-function,
especially for small values of $b$. As $b$ tends to infinity however, existing
bounds of $Q_{1}(a,b)$ yield to a closer representation of the function.

\section*{Acknowledgement}

The authors would like thank Toshiba Telecommunications Research Laboratory
and the EPSRC (grant EP/H500316/1) for their continued support.

\bibliographystyle{IEEEtran}
\bibliography{IEEEabrv,library}

\begin{thebibliography}{10}
\providecommand{\url}[1]{#1}
\csname url@samestyle\endcsname
\providecommand{\newblock}{\relax}
\providecommand{\bibinfo}[2]{#2}
\providecommand{\BIBentrySTDinterwordspacing}{\spaceskip=0pt\relax}
\providecommand{\BIBentryALTinterwordstretchfactor}{4}
\providecommand{\BIBentryALTinterwordspacing}{\spaceskip=\fontdimen2\font plus
\BIBentryALTinterwordstretchfactor\fontdimen3\font minus
  \fontdimen4\font\relax}
\providecommand{\BIBforeignlanguage}[2]{{%
\expandafter\ifx\csname l@#1\endcsname\relax
\typeout{** WARNING: IEEEtran.bst: No hyphenation pattern has been}%
\typeout{** loaded for the language `#1'. Using the pattern for}%
\typeout{** the default language instead.}%
\else
\language=\csname l@#1\endcsname
\fi
#2}}
\providecommand{\BIBdecl}{\relax}
\BIBdecl

\bibitem{Simon2003}
M.~K. Simon and M.-S. Alouini, ``Some {N}ew {R}esults for {I}ntegrals
  {I}nvolving the {G}eneralized {M}arcum {Q}- {F}unction and their
  {A}pplication to {P}erformance {E}valuation over {F}ading {C}hannels,''
  \emph{{IEEE} Trans. Wireless Commun.}, vol.~2, no.~4, pp. 611--615, 2003.

\bibitem{Odriscoll2009a}
C.~O'driscoll and C.~Murphy, ``A {S}implified {E}xpression for the
  {P}robability of {E}rror for {B}inary {M}ultichannel {C}ommunications,''
  \emph{{IEEE} Trans. Commun.}, vol.~57, no.~1, pp. 32--35, 2009.

\bibitem{Fu2011}
H.~Fu and P.-Y. Kam, ``Exponential-{T}ype {B}ounds on the {F}irst-{O}rder
  {M}arcum {Q}-{F}unction,'' in \emph{Proc. IEEE Global Telecommunications
  Conf. (GLOBECOM 2011)}, 2011, pp. 1--5.

\bibitem{Zhao2008}
X.~Zhao, D.~Gong, and Y.~Li, ``Tight {G}eometric {B}ound for {M}arcum
  {Q}-{F}unction,'' \emph{Electronics Letters}, vol.~44, no.~5, pp. 340--341,
  2008.

\bibitem{Simon2000}
M.~K. Simon and M.-S. Alouini, ``Exponential-{T}ype {B}ounds on the
  {G}eneralized {M}arcum {Q}-{F}unction with {A}pplication to {E}rror
  {P}robability {A}nalysis {O}ver {F}ading {C}hannels,'' \emph{{IEEE} Trans.
  Commun.}, vol.~48, no.~3, pp. 359--366, 2000.

\bibitem{Ding2008}
N.~Ding and H.~Zhang, ``A {F}lexible {M}ethod to {A}pproximate {M}arcum
  {Q}-{F}unction {B}ased on {G}eometric {W}ay of {T}hinking,'' in \emph{Proc.
  3rd Int. Symp. Communications, Control and Signal Processing ISCCSP 2008},
  2008, pp. 1351--1356.

\bibitem{Sofotasios2010a}
P.~C. Sofotasios and S.~Freear, ``Novel {E}xpressions for the {M}arcum and
  {O}ne {D}imensional {Q}-{F}unctions,'' in \emph{Proc. 7th Int Wireless
  Communication Systems (ISWCS) Symp}, 2010, pp. 736--740.

\bibitem{Bjorck1996}
\BIBentryALTinterwordspacing
B.~Ake, \emph{Numerical {M}ethods for {L}east {S}quares {P}roblems}.\hskip 1em
  plus 0.5em minus 0.4em\relax Society for Industrial and Applied Mathematics,
  1996. [Online]. Available:
  \url{http://epubs.siam.org/doi/abs/10.1137/1.9781611971484}
\BIBentrySTDinterwordspacing

\bibitem{Luxmoore1979}
\BIBentryALTinterwordspacing
A.~R. Luxmoore, ``Statistical {M}ethods for {E}ngineers and {S}cientists,''
  \emph{International Journal for Numerical Methods in Engineering}, vol.~14,
  no.~2, pp. 313--313, 1979. [Online]. Available:
  \url{http://dx.doi.org/10.1002/nme.1620140217}
\BIBentrySTDinterwordspacing

\bibitem{Proakis2000}
J.~Proakis, \emph{Digital {C}ommunications}, 4th~ed.\hskip 1em plus 0.5em minus
  0.4em\relax McGraw-Hill Higher Education, Sept 2000, iSBN-13: 978-0072321111.

\bibitem{Miorandi2008}
D.~Miorandi, ``The {I}mpact of {C}hannel {R}andomness on {C}overage and
  {C}onnectivity of {A}d {H}oc and {S}ensor {N}etworks,'' \emph{{IEEE} Trans.
  Wireless Commun.}, vol.~7, no.~3, pp. 1062--1072, 2008.

\bibitem{Coon2012a}
\BIBentryALTinterwordspacing
J.~Coon, C.~P. Dettmann, and O.~Georgiou, ``\BIBforeignlanguage{English}{Full
  {C}onnectivity: {C}orners, {E}dges and {F}aces},''
  \emph{\BIBforeignlanguage{English}{Journal of Statistical Physics}}, vol.
  147, pp. 758--778, 2012. [Online]. Available:
  \url{http://dx.doi.org/10.1007/s10955-012-0493-y}
\BIBentrySTDinterwordspacing

\bibitem{Li2010}
R.~Li, P.~Y. Kam, and H.~Fu, ``New {R}epresentations and {B}ounds for the
  {G}eneralized {M}arcum {Q}-{F}unction via a {G}eometric {A}pproach, and an
  {A}pplication,'' \emph{{IEEE} Trans. Commun.}, vol.~58, no.~1, pp. 157--169,
  2010.

\bibitem{Pent1968}
M.~Pent, ``Orthogonal polynomial approach for the marcum qfunction numerical
  computation,'' \emph{Electronics Letters}, vol.~4, no.~25, pp. 563 --564, 13
  1968.

\end{thebibliography}
\bigskip
\end{document}